**Phonons, Q-dependent Kondo spin fluctuations, and 4*f*/phonon resonance in YbAl$_3$**

Andrew D. Christianson, Victor R. Fanelli, and Lucas Lindsay, Oak Ridge National Laboratory, Oak Ridge, TN 37831.
Sai Mu, University of California, Santa Barbara, CA 93106
Marein C. Rahn, Technical University of Dresden, 01062 Dresden, Germany
Daniel G. Mazzone, Brookhaven National Laboratory, Upton, NY 11973
Ayman H. Said, Argonne National Laboratory, Lemont, IL 60439
Filip Ronning, Eric D. Bauer, and Jon M. Lawrence, Los Alamos National Laboratory, Los Alamos, NM 87545

**Abstract**

The intermediate valence (IV) compound YbAl$_3$ exhibits nonintegral valence (Yb $4f^{14-n_f}(5d6s)^z$ where $z = 2+n_f = 2.75$) in a moderately heavy (m* ~ 20-30m$_e$) ground state with a large Kondo temperature (T$_K$ ~ 500-600K). We have measured the magnetic fluctuations and the phonon spectra on single crystals of this material by time-of-flight inelastic neutron scattering (INS) and inelastic x-ray scattering (IXS). We find that at low temperature, the Kondo-scale spin fluctuations have a momentum (**Q**) dependence similar to that seen recently in the IV compound CePd$_3$ and which can be attributed to particle-hole excitations in a coherent itinerant 4*f* correlated ground state. The **Q**-dependence disappears as the temperature is raised towards room temperature and the 4*f* electron band states become increasingly incoherent. The measured phonons can be described adequately by a calculation based on standard DFT+*U* density functional theory, without recourse to considering 4*f* correlations dynamically. A low temperature magnetic peak observed in the neutron scattering at ~ 30meV shows dispersion identical to an optic phonon branch. This 4*f*/phonon resonance disappears for T ≥ 150K. The phonons appear to remain unaffected by the resonance. We discuss several possibilities for the origin of this unusual excitation, including the idea that it arises from the large amplitude beating of the light Al atoms against the heavy Yb atoms, resulting in a dynamic 4*f*/3*p* hybridization.

## Introduction

Intermediate valence (IV) materials[1] are rare earth intermetallic compounds, including $CePd_3$, $CeSn_3$, and $YbAl_3$, in which a localized $4f$ state hybridizes with continuum conduction electrons in the presence of strong on-site Coulomb interactions that prevent multiple occupancy of the $4f$ state. The $4f$ occupancy is fractional, e.g. for $CePd_3$, the occupation is $n_f \sim 0.75$, representing 75% occupancy of the $4f^1(5d6s)^3$ state and 25% occupancy of the $4f^0(5d6s)^4$ state and a resulting valence $z = 4-n_f \sim 3.25$.[2] The correlations lead to a nonmagnetic moderately heavy fermion ground state with an enhanced effective mass m* in the range 20-50$m_e$, which gives an enhanced specific heat $\gamma = C/T \propto m^*$ in the range 50mJ/mol-$K^2$ and an enhanced Pauli susceptibility $\chi_p(T=0)$, also proportional to m*. The correlated electron physics is essentially that of the Kondo/Anderson lattice, and the large spin fluctuation (Kondo) energy $E_K = k_B T_K$ with $T_K \sim$ 500K can be observed as a broad peak at $E_K$ in the dynamic susceptibility $\chi''(E)$. The latter can be measured via inelastic neutron scattering. It has been found that the Anderson Impurity Model (AIM) describes much of the physics over a broad range of energy and temperature.[2,3] The behavior of the ground state of the IV lattice, however, is "coherent," i.e. it exhibits properties indicating that the $4f$ electrons are itinerant. For example, the neutron scattering in single-crystalline $CePd_3$ has been shown[4] recently to exhibit a dependence on momentum and energy transfer (**Q**,$\Delta E$) expected for particle-hole scattering in strongly renormalized $4f$ bands. The data was shown to fit quantitatively to the results of a calculation using density functional theory (DFT) where the correlations were treated by dynamic mean field theory (DMFT).

In these compounds the Kondo temperature is considerably larger than the crystal field splitting, so that the latter can be ignored. Because the nonmagnetic state does not lie in close proximity to a quantum critical point (QCP) for transition to a magnetically ordered state, the complications of QCP physics (e.g. strong antiferromagnetic spin correlations) can also be ignored. The IV compounds hence represent a comparatively simple version of the correlated electron problem and can give insight into the behavior of the larger set of correlated materials.

The IV compound $YbAl_3$ has a ground state $4f$ occupancy $n_f \sim 0.75$, representing a state with 75% $4f^{13}(5d6s)^3$ and 25% $4f^{14}(5d6s)^2$ and a resulting valence $z = 2+n_f = 2.75$. (Note that for Yb, $n_f$ is the number of holes in the $4f^{14}$ shell.) The specific heat coefficient is of order 40mJ/mol-$K^2$ and de Haas van Alphen (dHvA) measurements[5] display orbits with effective masses of order 20$m_e$. The specific heat coefficient, magnetic susceptibility, and $4f$ occupation number can be fit semi-quantitatively by the AIM where the Kondo temperature is of order 500-600K.[6] A quadratic temperature dependence of the resistivity is seen below a Fermi liquid temperature $T_{FL} \sim$ 50K. The optical conductivity shows a broad peak in the mid-infrared ($E_{max} \sim 0.25$eV) which is characteristic of heavy fermion compounds; at lower energy a narrow Drude tail is separated from the mid-infrared by a deep minimum in the conductivity at 30meV.[7]

Early inelastic neutron scattering measurements of polycrystalline samples of $YbAl_3$ showed three features.[8] First, broad phonon peaks were observed near 12, 21, and 32meV. Second, the high temperature magnetic spectrum was observed to be quasielastic $\chi''(\Delta E) \sim \Delta E\, \Gamma/(\Delta E^2 + \Gamma^2)$ with a characteristic energy $\Gamma$ equal to 35meV at 200K and increasing to 45meV at 100K. Below 100K, the scattering becomes inelastic ($\chi''(\Delta E) \sim \Delta E\, \Gamma/((\Delta E-E_0)^2 + \Gamma^2)$); at 5K a maximum was seen near $E_0 \sim$ 45meV. We note that a similar transfer of magnetic spectral weight from a low energy quasielastic peak at high temperature to a higher energy inelastic peak at low temperature is seen in CePd$_3$.[2,4] A third feature – a narrow peak near 30meV – was also observed in YbAl$_3$ at low temperature. The intensity in this peak disappeared rapidly with increasing temperature. Although the peak was seen to be coincident with an optic phonon, the fact that it did not exhibit the characteristic increase with temperature expected for phonons led to an alternate explanation: that the peak was due to scattering across an indirect hybridization gap.[9] The peak was found to disappear rapidly on alloying with Lu, and hence required lattice coherence.[10] An initial study[11] of a single crystal sample found that the 30meV peak exhibited a very weak dependence on momentum transfer **Q**; this led to the interpretation that the state represented a spatially localized excitation in a hybridization gap. That study also found that the Kondo-scale scattering, which at low temperature peaked near 50meV, displays a marked **Q**-dependence with high intensity in the vicinity of the R point.

Due to limitations of neutron intensity, the kinematics in the earlier single-crystal experiment were quite restricted. Recent advances in neutron scattering instrumentation have allowed rapid data collection where the sample can be rotated with respect to the incoming beam and the data interpolated to give essentially full coverage of four-dimensional (H,K,L, ΔE) space. In this paper, we employ these advances to report a new study of the magnetic and vibrational properties of single-crystal YbAl$_3$ via neutron scattering. Inelastic x-ray scattering (IXS) is used as a complementary probe that isolates the non-magnetic character of the excitations. We show that the Kondo-scale scattering is **Q**-dependent in a manner similar to that of CePd$_3$, and hence represents particle-hole scattering in the ground state correlated f-bands. We have also determined the phonon spectra, which compare favorably to results of a calculation employing a standard DFT+$U$ band-theoretic approach. Finally, we show that the 30meV excitation, far from lacking **Q**-dependence as reported earlier, has dispersion identical to that of a particular optic phonon branch. Since this excitation is clearly magnetic, we speculate that it arises from a 4$f$/phonon resonance, and we discuss the conditions required for the resonance to occur.

E**xperimental and Theoretical Details**

Single crystals of YbAl$_3$ were grown in Al flux as reported earlier.[12] Six flux-grown crystals, each with a mass of approximately one gram, were co-aligned to within 0.3° with the [1,-1,0] direction vertical. Neutron scattering measurements were performed on the ARCS beamline at the Spallation Neutron Source (SNS) at Oak Ridge National Laboratory. Three incident energies were used (60, 80, and 120meV); these are indicated in the body and/or caption of each figure. The measurements were carried out at several temperatures between 5 and 300K. The angle between the incident neutron beam and the sample axes was varied over a 90º range in 1º steps. The data was interpolated using the Dave software[13] for polycrystalline averages and using the Horace software[14] for analyzing over the four dimensions of momentum and energy transfer. We note that in this paper, the momentum transfer **Q**=2π/a$_0$(H,K,L) will be given as Miller indices (H,K,L), i.e. in reduced lattice units, for which the Γ-point is at (0,0,0) and equivalent Q-vectors, X is at (1/2,0,0), M is at (1/2,1/2,0) and R is at (1/2,1/2,1/2).

To characterize lattice fluctuations in the AuCu$_3$-type (Pm$\overline{3}$m) structure of YbAl$_3$ in the charge channel, inelastic x-ray scattering (IXS) measurements were performed at the HERIX spectrometer in Sector 30 of the Advanced Photon Source (APS), Argonne National Lab.[15]

A single crystal of YbAl$_3$ was cut and polished to 50μm thickness. This allowed for 30% x-ray transmission at a photon energy of 23.71keV (λ=0.522Å), which was used throughout the experiment. The scattering from the sample was investigated with momentum transfers in the (HK0) plane of reciprocal space. To provide sensitivity to different phonon polarizations, spectra were recorded for each phonon momentum $q_{ph}$ at a number of equivalent momentum transfers $Q=q_{(HK0)}+q_{ph}$ with Q almost parallel to either the [1,0,0] or the [1,1,0] direction of reciprocal space. Spectra were recorded at 300K and 10K with energy transfers up to 40meV and a combined energy resolution of dE≈ 1.6meV. To determine the energies and widths of phonon excitations, peaks in the spectra were fitted with the lineshape of a damped harmonic oscillator (DHO)

$$S_{DHO} \propto (n(E)+1)\gamma E/((E_q^2+\gamma^2-E^2)^2+4\gamma^2 E^2) ,$$

where E is the energy transfer, n(E) is the Bose occupation factor, $E_q$ is the phonon energy, and γ is the damping constant.[16,17]

The electronic band structures and the phonon dispersions of YbAl$_3$ were calculated via density functional theory (DFT). We used the Vienna *ab-initio* simulation package (VASP) within the DFT+*U* generalized gradient approximation (GGA) formalism and the projector augmented wave (PAW) method.[18] PAW pseudopotentials correspond to the valence electron configuration $(5s)^2(5p)^6(4f)^{14}(5d)^0(6s)^2$ for Yb and $(3s)^2(3p)^1$ for Al. The plane wave kinetic energy cutoff was set to 450 eV. Variation of the Coulomb correlation energy *U* for the 4*f* shells of the Yb atoms was included in a spherically averaged scheme.[19] The band structure and phonon spectra were calculated with and without spin-orbit coupling (SOC).

Figure 1 demonstrates the dependence of the calculated electronic bands on the Coulomb correlation energy and spin orbit coupling. The effect on the electronic bands of increasing the Coulomb correlation energy in the range *U*=2-6eV was to shift the flat occupied f-band to lower energy while the effect of the spin orbit interaction was to split the flat occupied f-band. The states above the Fermi level were unaffected by either interaction. We note that the 4*f* bands in our calculation are nearly fully occupied (4f$^{14}$(5d6s)$^2$), which has been the case for all previous band calculations of YbAl$_3$.[20] Some hybridization of the otherwise divalent 4$f^{14}$ orbitals was found after self-consistent solution of the charge density giving estimated calculated *f* charge of 13.7. The actual material has an f-count 13.3, which means the Yb is nearly trivalent rather than nearly divalent. Because of this fact, the earlier DFT studies of the band structure had to include an additional 0.2Ry shift of the f-level to reproduce the dHvA frequencies and hence obtain the correct Fermi volume. It is not clear why the Fermi volume is given incorrectly by the DFT calculations. Despite this, we note that calculations including SOC and *U*=6 eV reproduce *f* states with binding energies of approximately -1 eV and -2.5 eV, consistent with previous angle-resolved photoemission spectroscopy (ARPES) measurements though shallower than previous calculations using an all-electron full-potential linear augmented plane wave method.[21]

The phonon spectra were essentially unaffected by either the Coulomb energy or the spin-orbit interaction. Since the divalent lattice constant found in the theory is larger than that measured experimentally for the intermediate valent compound, we performed the calculation with the lattice constant constrained to the experimental value. While the 4*f* occupancy remains close to $f^{14}$ under this change, the phonon spectra are shifted to higher energies, in better agreement (although still 8% low) compared with the experimentally measured values. Underestimation of phonon frequencies is a known characteristic of GGA calculations, which tend to underbind the atoms giving lattice constants typically larger than measured values.[22] (Relevant VASP input and output files for replicating the electronic and vibrational band structures with the experimental lattice parameter, $U$=6eV, and including SOC are provided as supplemental materials.[23])

**The Phonons**

The calculated phonon spectra are shown in Fig. 2. It is clear that variation of the Coulomb correlation energy and the spin-orbit splitting have little effect on the spectra, whereas constraining the lattice constant of the calculation to the experimental value shifts the spectra to higher energy. Where there is overlap (along ΓR and ΓM) the calculated spectra for the relaxed lattice constant are in good agreement with those of an earlier calculation.[24] In Fig. 3 we plot the phonon structure factor at two locations in the vicinity of the Γ-point. The structure factor at the momentum transfer **Q**=(3,3,1) is very weak compared to that observed at (3,3,2). Apart from the difference in neutron scattering lengths of Al (3.45fm) and Yb (12.41fm), this is essentially because the optic modes at Γ primarily involve motions of the light Al atoms where the three Al polarizations within the unit cell cancel. Under these circumstances, the structure factor is similar to that of a **Q**=0 fcc antiferromagnet. On the other hand, for the acoustic modes, the Yb and Al atoms move in the same direction, so that the all odd (all-even) scattering condition of the fcc lattice is approximately valid.

In Fig. 4 we present energy – momentum space color maps of the experimentally measured neutron spectrum showing the phonons along three directions: ΓX, ΓM, and ΓR. The measurements were performed at 150K using an initial neutron energy $E_i$ = 60meV which provides better energy resolution than in measurements at $E_i$ = 120meV. We chose sufficiently large values of Q (= |**Q**|) that the phonons dominate over the background and the magnetic scattering. We plot in two different Brillouin zones because, as mentioned, the phonon structure factor for certain branches vanishes for certain (H,K,L). In agreement with the calculation, we observe acoustic modes and two regions of optic modes. To quantitatively reproduce the measured phonon energies, we used the calculation with the lattice constant constrained to the experimental value and with the calculated energies uniformly multiplied by an additional factor of 1.08. With this scaling, the overall agreement between experiment and theory is quite good, with a few exceptions such as at the R point for the optic modes near 30meV.

Figure 5 shows the dispersion of acoustic phonons at the Brillouin zone centers at 300K and 10K measured by inelastic x-ray scattering (IXS). The corresponding velocities were determined by fitting linear slopes (or, where appropriate, sine functions), as drawn in the figures. The elastic constants $C_{ij}$ of a cubic crystal are defined as

$$C_{11} = \rho V_L^2 [100]$$
$$C_{44} = \rho V_T^2 [100]$$
$$[(C_{11}+C_{12})/2] + C_{44} = \rho V_L^2 [110]$$
$$(C_{11}-C_{12})/2 = \rho V_T^2 [110]$$

where ρ is the mass density, and $V_L$[HK0] and $V_T$[HK0] denote the velocities of the longitudinal and transverse acoustic phonons, respectively, propagating along the [HK0] direction of reciprocal space. The elastic constants thus derived are summarized in Table 1.

Figure 6 exhibits the phonon dispersion at 300K as measured by IXS. To indicate the distribution of the scattered intensity, the IXS spectra measured at momentum transfers along the high-symmetry directions of the crystal are shown as color plots. Superimposed markers indicate the fitted positions and full-widths at half-maximum of fitted damped harmonic oscillator (DHO – see above) lineshapes. Blue and green markers correspond to features observed with an overall momentum transfer almost parallel to the [100] and [110] directions, respectively. The linear dispersion of acoustic phonons at the Brillouin zone centers is also indicated. The inelastic x-ray scattering measurements thus reveal a phonon dispersion in very good agreement with both the computational results and the neutron scattering results discussed above.

**Overview of the Magnetic Fluctuations: Polycrystalline Averaged Scattering**

For an overview of the low temperature magnetic neutron scattering, we plot in Fig. 7a the polycrystalline average of the single crystal scattering intensity at T=5K, measured with an incident energy $E_i$=120meV. There are three important regions of the data. Below 30meV, there are two peaks near 14 and 22meV whose scattering increases with momentum transfer Q = |**Q**| (Figs. 7a and c). These are due to phonons, the 14meV peak representing the van Hove singularity for the acoustic phonons and the 22meV peak involving an average over the intermediate optic phonon branch. (The oscillations in Fig. 7c arise from the variation with Q of the averaged phonon structure factor.) Above 35meV, there is a broad excitation that decreases in intensity as the momentum transfer increases (Fig. 7a). The scattering in this energy range tracks the Yb($3^+$) $4f$ form factor, as seen for ΔE=50meV in Fig. 7d, and hence represents magnetic scattering. We will refer to this as Kondo-scale scattering, as its energy (48meV) closely corresponds to the Kondo energy (500-600K) expected for this material. A third peak is observed near 32meV. This peak shows very little Q-dependence in the range $2 < Q < 6$Ang$^{-1}$ (Fig. 7a), but on closer examination (Fig. 7d) the scattering at 5K appears to initially decrease with Q and then to increase. At 100K, the scattering increases with Q in a manner that is essentially similar to that seen at 14meV, although substantially weaker, and hence represents the optic phonons in this energy range. On subtracting the Q-dependence at 100K from that at 5K, the result closely tracks the Yb($3^+$) $4f$ form factor.[25] Therefore, it appears that at 5K the 32meV peak arises from a combination of both phonon and magnetic scattering. The latter is a low temperature effect which has nearly vanished by 100K.

To determine the magnetic spectra, we subtracted the background and phonon scattering from the total measured scattering. Above 40meV, where the phonons are absent, the background can be determined from fits such as that shown in Fig. 7d for ΔE=50meV. We assume that the low temperature magnetic scattering is vanishingly small below 25meV, so the background can be determined as the Q=0 extrapolation of such data as in Fig. 7c for ΔE=14meV.[26] The resulting background is plotted as a solid black line in Fig. 7a. The phonons can be determined from the large-Q scattering peaks at 14, 22, and 32meV and extrapolated to smaller Q using the measured Q dependence. The phonons at higher temperature are scaled up from the low temperature value by the thermal factor n(ΔE)+1. The total fits are shown in Fig. 7a and the Q=0 magnetic scattering is shown in Fig. 7b.

At 300K, the magnetic scattering is approximately quasielastic. As the temperature is lowered, magnetic spectral weight shifts to higher energy. At low temperature the scattering can be fit by an inelastic Lorentzian Kondo peak centered near 50meV plus an additional magnetic peak at 32meV. If the inelastic Kondo peak is viewed as exhibiting a pseudogap below 30meV, then the 32meV excitation appears to reside near the gap edge. (We note the similarity of our results to those of the earlier study of polycrystalline $YbAl_3$.[8,9])

**Q-Resolved Kondo-Scale Magnetic Scattering**

Based on the observations[2,4] of $CePd_3$ the **Q**-resolved low temperature scattering in $YbAl_3$ is expected to possess (vector) **Q** dependence beyond that of the form factor. This is shown in Fig. 8 at T=5K ($E_i$=120meV) for energies ΔE >40meV above the phonon cutoff. In the upper panel (Fig. 8a), a slice of the data with axes HH (L=3/2) and ΔE exhibits columns of intensity, with stronger scattering near (1/2,1/2,3/2) and weaker scattering at (0,0,3/2). A slice in the (HH,L) plane at the peak energy of the Kondo scattering (48meV) shows spots of high intensity near (1/2,1/2,3/2) and weaker scattering at (0,0,3/2), (1/2,1/2,2), and (0,0,2) (Fig. 8b). The cuts (Fig. 8c) along HH at L=3/2 and 2 show that the 48meV excitation is strongest near (1/2,1/2,1/2), weaker at (0,0,1/2) and (1/2,1/2,0), and weaker still at (0,0,0).

To determine the magnetic scattering we subtract the background (determined as above) from I(ΔE) spectra at four high symmetry values of **Q**. At the small Q=|**Q**| of these spectra, phonon scattering is negligible. We divide the background-subtracted data by the 4*f* form factor to reduce the data to the first Brillouin zone. Figure 9 demonstrates that at low temperature, the scattering in the Kondo peak near 48meV decreases in the sequence I(1/2,1/2,1/2) > I(1/2,1/2,0) > I(1/2,0,0) > I(0,0,0) similar to what is seen in Fig. 8c. The scattering in the 32meV peak has a different sequence, with I(1/2,0,0) > I(0,0,0) > I(1/2,1/2,0) > I(1/2,1/2,1/2). At 100K, the 32meV peak is quite weak and the Kondo-scale scattering is weaker than at 5K, but it appears to maintain the same sequence of intensity with **Q**. At 300K, the 32meV peak is absent, while the Kondo-scale scattering is independent of **Q** and is approximately quasielastic with the same width parameter (Γ=18meV) as seen in Fig. 7b for the polycrystalline averaged scattering.

**The 4*f* /phonon resonance**

The polycrystalline scattering shown above demonstrates that a magnetic peak near 32meV that appears only below about 150K is coincident with optic phonons at the same energy. To examine this in more detail, we plot in Fig. 10 the total spectra (background, phonon, and magnetic intensity) for the **Q**-resolved scattering at Γ, X, M, and R, at both 5K and 150K. In these plots, the solid line is the background, essentially the same as for polycrystalline scattering. The dashed line represents the broad magnetic scattering seen at 150K (similar to the 100K scattering seen in Fig. 7b). The peaks near 22meV arise from the optic phonon branches at this energy. It can be seen that these peaks do not change appreciably with temperature, due to the fact that the thermal (Bose) factor remains close to unity at this energy at 150K. Phonon peaks arising from the uppermost optic phonon branches can be seen at (1/2,1/2,3/2) and (3/2,3/2,0); the amplitudes of these peaks do not change between 5 and 150K. No phonons are seen at (1,1,1) and (1,1,3/2), both because of the small value of Q and due to the vanishing of the phonon structure factor for this zone. The magnetic scattering at 5K can then be determined by subtracting the background (solid line) and the phonon observed at 150K from the total scattering. The result is shown in Fig. 11a, which shows a similar trend as in Fig. 9a, but with better energy resolution. Essentially, a magnetic scattering peak emerges near 31meV at low temperature, which is most intense at Γ and X, weak at M, and negligible at R. (In earlier work,[11] we observed that the mode is weakest at R.) The data at the R-point suggests that this resonance may also occur near 35meV. At higher energy, the peak merges into the Kondo-scale scattering seen in Figs. 7b and 9a. In Fig. 11b we show how the peak at Γ disappears as the temperature is raised.

An important issue is the relationship between the magnetic peak and the optic phonon with which it coincides. In Fig. 12, for which the data is taken near the (1,1,1) Γ-point where the phonon structure factor vanishes, no scattering is seen at 150K. It can be seen that at 5K the leading edge of the resonance tracks the lower optic phonon dispersion near 32meV. At 150K, the resonance has disappeared. The spectra of Fig. 13 also show that the energy of the magnetic peak tracks the energy of the phonon peak, except at Γ where the magnetic peak occurs at roughly 1meV lower energy. The result of this analysis is shown as solid points in Fig. 4a. It thus appears that the magnetic peak involves a resonance with this optic phonon branch.

In Figure 14a we fit the background-subtracted scattering to the sum of two phonons (as observed at 150K) and a magnetic line. We have modeled the latter with a Fano lineshape, which provides a good fit for the magnetic resonance at the Γ-point. However, the data of Fig. 11a suggest that resonance may also occur for the 35meV optic phonon branch. This seems especially clear at the R-point. Hence, even though the magnetic spectrum at Γ is asymmetric, as for the Fano case, the spectrum may reflect this additional resonance rather than a Fano process. Otherwise said, the good fit to the Fano lineshape may be fortuitous. The scattering at 5K at equivalent but larger Q=|**Q**| is shown in Fig. 14b and c. Given that the thermal (Bose) factor does not change the phonon amplitude between 5 and 150K for $\Delta E > 30$meV, it can be seen that the increase in intensity at 5K relative to the 150K data is essentially the magnetic scattering reduced by the 4$f$ form factor. These data thus not only confirm that the resonant peak is magnetic, but also that the resonance has no effect on the energy or amplitude of the phonon.

Whereas for neutron scattering, the anomalous magnetic scattering contribution near 32meV overlaps with the optical phonon scattering, inelastic x-ray scattering (IXS) measures only the lattice contribution. Fig. 15 shows that, aside from the expected thermal expansion, softening, and phonon occupation, there is no anomalous variation between IXS datasets at 300K and 10K at the X-point for either case where **Q** is parallel or not parallel to **q**$_{ph}$. The insets show this on an expanded scale for the 32-35meV phonon branch. This confirms that no renormalization of the phonon energies, or relative spectral weight of the phonon branches is observed within the sensitivity of the present measurement. Otherwise said, although an enhancement of magnetic excitations in YbAl$_3$ is associated with optical phonons in the 32-35meV range, the lattice dynamics are not significantly affected by this phenomenon.

**Discussion**

Both neutron and x-ray spectra appear to be in very good agreement with the computational phonon dispersions for the case that the lattice constant in the DFT+$U$ calculation is constrained to the experimental value and the calculated energies are multiplied by an additional factor of 1.08. Since the calculations were performed using static DFT+$U$ band theory and resulted in a nearly integral valent compound, a possible reason for this underestimate of the phonon energies is that *dynamic* correlated hybridization (e.g. DFT+DMFT) would lead to a stronger binding energy and higher energy phonons in the intermediate valent material. In this regard, we note that the Kondo condensation energy (the energy lowering due to correlated hybridization) is large – $k_B T_K \sim$ 50meV. However, it is well-known that GGA calculations tend to underbind the atoms. This results in larger lattice constants and lower phonon energies and may justify the correction by 8% in and of itself. Said otherwise, it is to be expected that a DFT-based calculation including only static DFT+$U$ correlation effects will correctly obtain most of the details of the phonon spectra to within 10%, and this appears to be the case for YbAl$_3$.

The **Q**-dependence of the Kondo-scale scattering in YbAl$_3$ is qualitatively very similar to what was observed[4] in CePd$_3$, where columns of alternating strong and weak intensity similar to that shown in Fig. 8a were observed in 2D (**Q**,$\Delta$E) slices, bright spots similar to Fig. 8b were seen in 2D ($Q_1$,$Q_2$) slices at fixed $\Delta$E, and peaks similar to those of Fig. 8c were seen in cuts along various **Q**-directions at fixed energy. A difference between the two compounds is that the greatest intensity in the Kondo-scale peak occurs at different values of **Q**: near (1/2,1/2,1/2) for YbAl$_3$ and at (1/2,1/2,0) for CePd$_3$. In the earlier single-crystal study of YbAl$_3$, we found a peak for energy transfer 50meV at a location (1/2,1/2,1.3). The bright spots seen in Fig. 8b do indeed appear to be offset from the (1/2,1/2,1/2) point, occurring near (0.35,0.35,0.35). This is confirmed by the splitting of the peak seen in Fig. 8c. For CePd$_3$, we showed that the large intensity at (1/2,1/2,0) could be understood qualitatively as arising from transitions between a flat occupied band at $\Gamma$ and a flat unoccupied band at M. The theoretical calculation of $\chi''$(**Q**,$\Delta$E) that was successful for CePd$_3$ required evaluating a Lindhard function over the renormalized band structure determined via DFT+DMFT, with vertex corrections playing an important role in the final shape and intensity of the scattering. Apparently the location of the maximum intensity in YbAl$_3$ reflects a different renormalized band structure. Depending on the details, the high intensity scattering need not occur at high symmetry points such as R, which appears to be the case for YbAl$_3$.

The magnetic scattering at 100K is weaker and broader, but appears to continue to exhibit the same overall **Q**-dependence as at 5K in YbAl$_3$ (Fig. 9b). Hence the renormalized correlated band structure is still intact at 100K, but it experiences $\Delta$E- and Q-broadening as the scattering rate increases. As the temperature is raised to room temperature, the magnetic spectra become independent of **Q**. Furthermore, the spectral weight of the spin fluctuations shifts to lower energy as the temperature is raised until the scattering becomes quasielastic. This is believed to reflect the crossover from low temperature coherent to high temperature incoherent f-bands,[4] and is an important feature of correlated electron systems in general.[27]

The earliest explanation[8,9] for the 30meV magnetic excitation was that it represents scattering across an indirect gap. The latter occurs in the band structure of the Anderson lattice, where a sharp *f*-level crosses a parabolic band. The indirect gap then opens between the zone center and the zone boundary; the direct $\mathbf{Q} = 0$ scattering is pushed to much higher energies. This oversimplified band structure can be ruled out for YbAl$_3$, since the $\mathbf{Q} = 0$ (Γ) scattering occurs on a similar energy scale as the (1/2,1/2,1/2) scattering. On the other hand, the actual bands of YbAl$_3$ may include indirect band gaps. It therefore remains a possibility that the 30meV excitation arises from band-edge scattering. While a nearly-free electron model[20] of the bands in YbAl$_3$ indicated that no such gaps exist, it is plausible that a band theory which correctly incorporates dynamic correlated hybridization might prove otherwise. The very deep minimum in the optical conductivity[7] of YbAl$_3$ is suggestive of the existence of such a gap. Indeed, our earlier speculation[11] that the excitation represented a local state in a hybridization gap was based on this possibility.

However, our observation that this feature disperses and that it tracks the dispersion of the lowest optic phonon band near 30meV strongly suggests that the excitation arises from a 4*f*/phonon resonance. While it is unclear whether the Fano lineshape seen in Fig. 14 at the Γ-point is a fortuitous or a significant feature of the scattering, we note that recent Raman scattering measurements[28] of Nd(O,F)BiS$_2$ and optical conductivity measurements[29] of TaAs have proposed that the Fano lineshape arises from a resonance between a phonon mode and the electrons.

It is well known that the IV compounds possess strong electro-elastic coupling.[30] This leads to large thermal expansions and Grüneisen parameters, to isomorphic phase transitions as in γ-Ce where a large (5%) change in lattice constant occurs, and to specialized electron/phonon coupled modes.[31] There are several ways this might lead to a 4*f*/phonon resonance. The first would arise from 4*f*/phonon coupling of the form proposed recently for α-uranium.[32] This model adds a term
$$\Sigma \{\lambda_q [a^+_q + a^-_{-q}] (f^+_{k+q} f_k) \}$$
to the Anderson lattice which represents scattering of the 4*f* electrons by the phonons. Here the a and f symbols represent creation and annihilation operators for the phonons and f electrons. When the edge of the hybridization gap equals the optic phonon energy, the 4*f* electron and the phonon hybridize and a resonance occurs. The deep minimum in the optical conductivity at 30meV and the transfer of spectral weight of the spin fluctuations from low energy at high temperature into an inelastic peak at low temperature with a leading edge near 30meV both are suggestive of this scenario. (In CePd$_3$, the energy scale of the spin fluctuations is larger, and the phonon energies are smaller, so the condition that the hybridization gap edge coincides with the optic phonon does not occur and no resonance is observed.)

In an alternative model,[33] the phonons drive the Al atoms towards the Yb 4*f* atoms, and thereby cause the hybridization to oscillate. This dynamic hybridization can be represented by a 4*f*/phonon coupling term of the form
$$\Sigma \{V_{cf} + \lambda [a^+ + a^-]\} (f^+c + c^+f).$$
Here, the a, f, and c symbols represent the creation and annihilation operators for the phonons, *f* electrons, and conduction electrons, respectively, and the term $V_{cf}$ ($f^+c + c^+f$) is the usual 4*f*/conduction hybridization term in the Anderson lattice. This has the attractive feature that the 30meV optic phonons are known to primarily entail motion of the Al atoms, which are light and have large oscillation amplitudes, with the heavy Yb atoms essentially motionless. The 4*f*/conduction hybridization in YbAl$_3$ primarily involves hybridization between the 4*f* and the Al

3p electrons. As the optic vibrations cause the light Al atoms to periodically move towards the heavy Yb 4f atoms, the hybridization oscillates at a frequency corresponding to the phonon energy. We note that there are indications that enhanced 4f scattering at an optic phonon energy also occurs in the compound YbCuAl.[34] This suggests that the resonance might occur more generally when the intermediate valence involves hybridization of the 4fs with Al 3p electrons. A special mode observed[35] in $SmB_6$ may represent a similar phenomenon where the boron atoms provide the light, large amplitude oscillations for a resonance with the Sm 4f electrons. This may also help explain why no such resonance occurs in $CePd_3$, where the hybridization involves the much heavier Pd atom.

Such a dynamic hybridization between the 4f and conduction states has been discussed many years ago as the origin of the "gap mode" seen in Raman scattering in $Sm_{1-x}RE_xS$ (RE = Dy, Gd, Pr,Y) alloys.[31] The authors of that study attribute a new magnetic excitation to the coupling of incoherently fluctuating 4f charge to lattice distortions. There are, however, very significant differences between the behavior of the 32meV magnetic excitation in $YbAl_3$ and what is seen in the SmS alloys. In particular, the magnetic mode in SmS alloys appears in the gap between the acoustic and optic phonons, exhibits no dispersion, and furthermore it is independent of temperature up to 300K. The above-mentioned mode seen in $SmB_6$ also occurs in the gap between the acoustic and optic phonons.

Another possibility concerns recent photoemission measurements[21] on $YbAl_3$ which show a flat occupied f-band along the Γ-X direction. As the temperature is lowered, this 4f level approaches the Fermi level $\epsilon_F$, and at approximately 100K, the 4f level energy $E_f$ is 30meV below $\epsilon_F$. As the temperature is lowered below 100K, the level approaches even more closely to the Fermi energy. Here, there would be a fairly simple condition: that a phonon excited by a neutron has enough energy to transfer into the 4f level. This condition would be valid below 100K. Not only is the temperature scale correct, but the fact that the largest 4f/phonon resonance that we observe occurs along the Γ-X direction adds credence to this idea. It is not clear whether this scenario would also lead to 4f/phonon hybridization.

Indeed, a problem in any model where the 4f electrons and phonons hybridize is that there should be a hybridization repulsion/splitting causing the phonon spectrum itself to be altered. This has been recently observed in the compound $CeAuAl_3$ where sharp Ce-4f crystal field levels hybridize with acoustic phonon modes, leading to a characteristic hybridized-mode dispersion with hybridization gaps. [36] As we show in Figs. 14 and 15, this does not occur in $YbAl_3$. There is no change in the energy or amplitude of the optic phonon; the 4f resonance simply tracks the phonon dispersion without altering it.

A second criticism of the scenarios above – with the possible exception of the model where the phonon drives oscillating Yb 4f/Al 3p hybridization – is that there is no obvious preference for the specific optic mode that the magnetic resonance appears to track. We show the pattern of vibration of two optic modes near 32meV, one at the Γ-point and one at the M-point in Fig. 16. Both involve very similar motions that simultaneously compress and stretch the Yb/Al bonds along the face diagonals, but the former leads to a strong resonance, while the resonance at the M-point is absent or weak (Fig. 11a). It is not clear to us in general which modes should lead to a stronger dynamic hybridization.

A final question concerns whether the resonance occurs in the Yb charge or spin channel. The optical conductivity[7] arises from Q=0 particle-hole excitations similar to those seen by neutron scattering on the Kondo scale at the Γ-point. The fact that the optical conductivity shows a deep minimum near 30meV, i.e. the energy of the resonance, suggests that the resonance does not involve excitation of particle-hole pairs. As mentioned above, the theory of Yang and Riseborough[32] shows that a resonance can arise from electron/phonon coupling when the optic phonon is coincident with a conduction electron gap edge. However, in $YbAl_3$ it is the leading edge of the *spin* fluctuation spectrum that coincides with the optic phonon. These facts suggest that the resonance may arise from coupling of the phonons to the Yb spin channel. Perhaps this type of coupling may not lead to any alteration of the phonons, as we have observed.

**Conclusion**

We have shown that the phonons in $YbAl_3$ are well-replicated by calculations incorporating a standard band-theoretic approach that treats the 4*f* correlations within the static DFT+*U* formalism. The Kondo-scale spin fluctuations in this material exhibit a transfer of spectral weight from a lower energy **Q**-independent quasielastic peak at room temperature to a higher energy **Q**-dependent inelastic scattering peak as the temperature is lowered. The low temperature **Q**-dependence can be qualitatively understood as arising from particle-hole scattering in coherent heavy 4*f* bands. The crossover to incoherent **Q**-independent scattering at higher temperature is an expected feature of correlated electron systems.

A low temperature magnetic excitation near 30meV follows the dispersion of a particular optic phonon branch. The fact that it disappears at higher temperature indicates that it is a feature of the coherent ground state. We give several speculations about this 4*f*/phonon resonance, including the idea that it involves a dynamic hybridization between the 4*f* electrons and the optic phonon. However, we do not understand why typical features of level-crossing hybridization (alteration of the phonon spectrum, including level repulsion and the onset of energy gaps) do not occur, nor do we understand what distinguishes the particular optic mode with which the 4*f* electron resonates. Determining the origin of this unusual mode will require additional theory and experiment.

Future work to improve our understanding of this material could include dynamic correlated band theoretic calculations, e.g. in the DFT+DMFT mode. Hopefully this will yield a band structure appropriate for an intermediate valent (z=2.75) system, as opposed to a divalent Yb band structure. The inclusion of the correlations might possibly also improve the agreement between the calculated and measured phonon energies. Experimental studies of the optic phonons and the 4*f*/phonon resonance at higher resolution are also in order. The resonance between the spin fluctuations and the phonon modes at 32 meV in this mixed valent material suggest that coherent excitations of these phonon modes by optical pump/probe techniques may induce novel metastable electronic behavior as observed in other correlated electron systems[37] potentially yielding further insight into the Kondo coherent state.[38]

**Acknowledgments**

Research at the Spallation Neutron Source at Oak Ridge National Laboratory was supported by the Scientific User Facilities Division, Office of Basic Energy Sciences, U.S. Department of energy (DOE). ADC was supported by the U.S. Department of Energy, Office of Science, Basic Energy Sciences, Materials Sciences and Engineering Division. During the early stages of this project ADC was supported by the U.S. Department of Energy, Office of Science, Basic Energy Sciences, Scientific User Facilities Division. Density functional theory calculations were supported by the U.S. Department of Energy, Office of Science, Basic Energy Sciences, Materials Sciences and Engineering Division.   Work at LANL and the phonon calculations were supported by the U.S. DOE, Basic Energy Sciences, Division of Materials Science and Engineering. This research used resources of the Advanced Photon Source, a U.S. Department of Energy (DOE) Office of Science User Facility operated for the DOE Office of Science by Argonne National Laboratory under Contract No. DE-AC02-06CH11357. D.G.M. acknowledges support from the Swiss National Science Foundation, Fellowship No. P2EZP2_175092, and the U.S. Department of Energy, Office of Science, Basic Energy Sciences, Materials Science and Engineering Division, under Contract No. DE-SC0012704. M.C.R. is grateful for a postdoctoral fellowship by the Humboldt Society. This research has been supported by the Deutsche Forschungsgemeinschaft through the SFB 1143 and the Würzburg-Dresden Cluster of Excellence EXC 2147 (ct.qmat).

|                          | 300K    | 10K     |
|--------------------------|---------|---------|
| $C_{11}$                 | 166(2)  | 193(4)  |
| $C_{44}$                 | 81(2)   | 86(5)   |
| $(C_{11}+C_{12})/2+C_{44}$ | 177(2)  | 190(6)  |
| $(C_{11}-C_{12})/2$      | 78(4)   | 69(1)   |
| $\rightarrow C_{12}$     | 18(8)   | 35(12)  |

Table 1: Elastic constants in GPa of YbAl$_3$ derived from the fits of acoustic phonon velocities shown in Fig. 5.

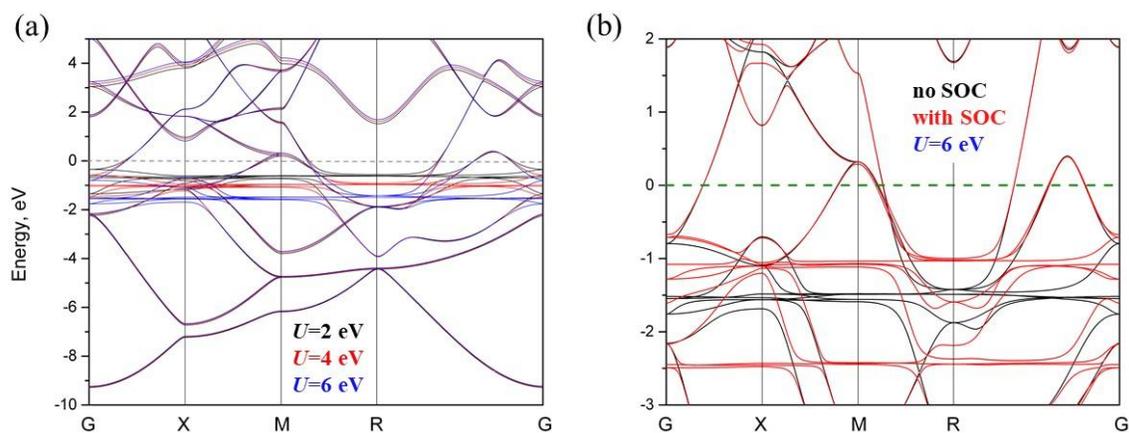

Figure 1: Calculated electronic band structure of YbAl$_3$ demonstrating the effects of (a) Coulomb correlation ($U$) and (b) spin-orbit coupling (SOC). Unlike the phonons, the electronic bands are sensitive to $U$ and SOC: larger $U$ values drive the flat $f$ electronic states further from the Fermi level (0 eV), while SOC splits the 4$f$ bands.

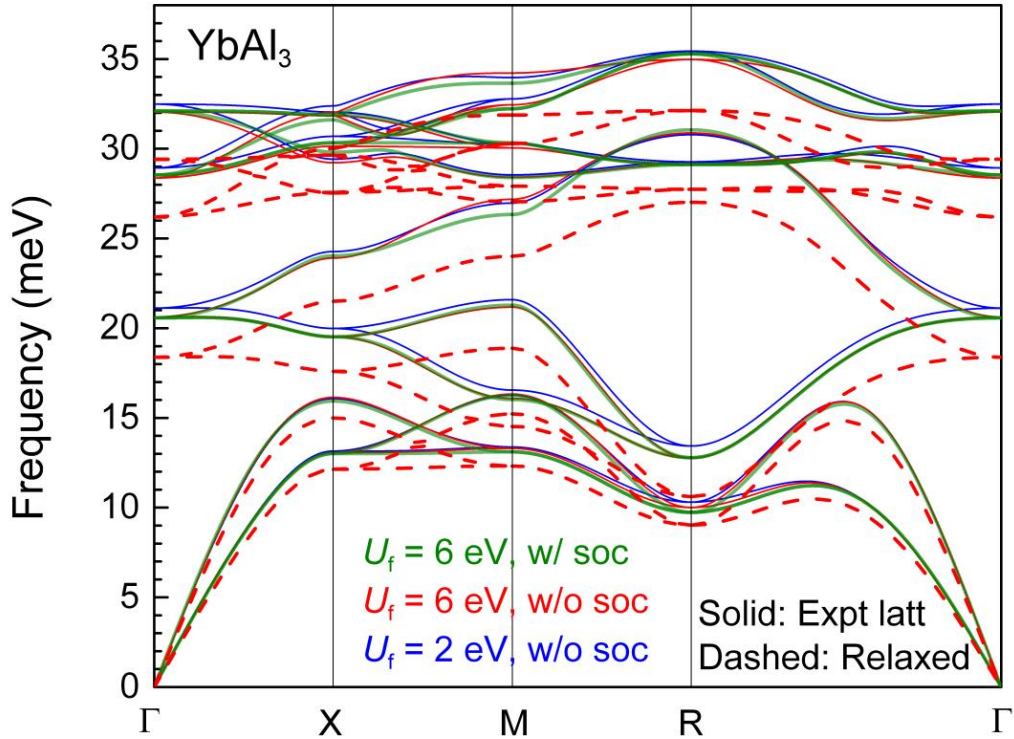

Figure 2: DFT calculated phonon spectra for YbAl$_3$ with varying lattice parameter, $U$ value, and spin-orbit coupling (SOC). Green solid curves correspond to $U=6$ eV with spin-orbit coupling and the experimental lattice constant ($a_{exp}$). Red solid and red dashed curves correspond to $U=6$ eV with no spin-orbit coupling using $a_{exp}$ and the GGA relaxed lattice constant, respectively. The GGA relaxed lattice constant is larger than the experiment value and thus gives softer phonons. Blue curves correspond $U=2$ eV with no spin-orbit coupling using $a_{exp}$. Note that spin-orbit interactions and varying $U$ in this range have little effect on the phonon dispersions.

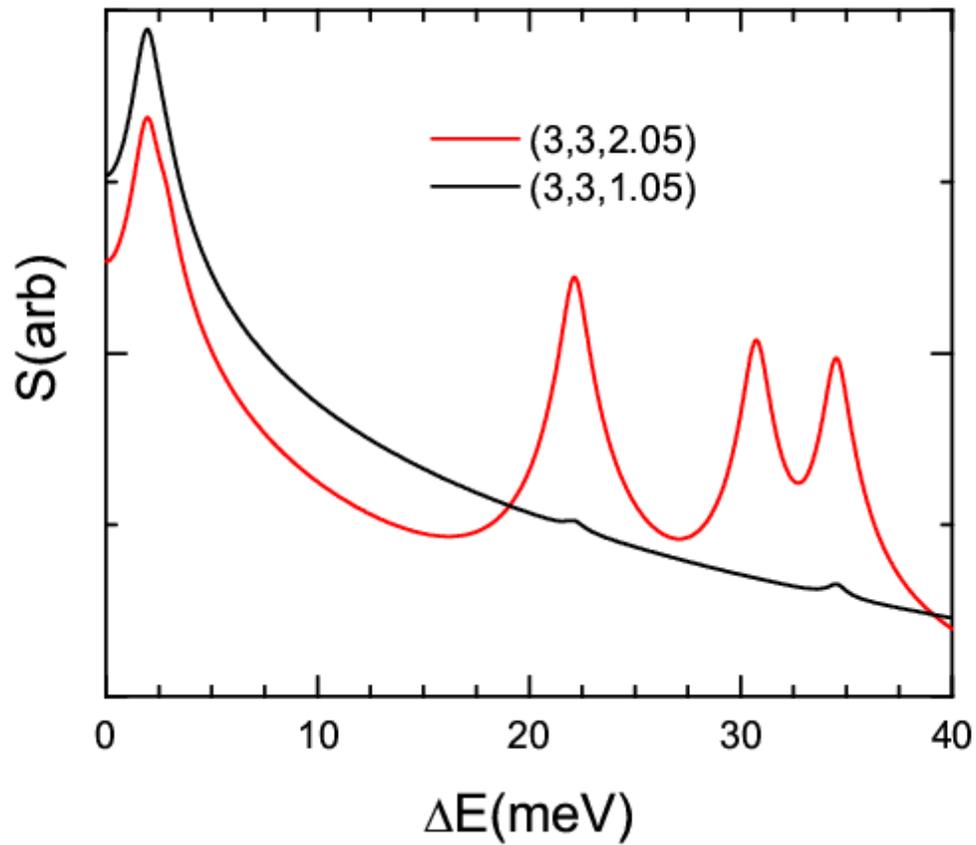

Fig. 3 The phonon structure factor for YbAl$_3$ calculated for two values of **Q**, one near the fcc-diffraction-allowed Γ-point (3,3,1) and one near (3,3,2) where fcc diffraction is not allowed. The data have been subjected to a Gaussian broadening. The very small structure factor near (3,3,1) reflects the fact that the aluminum atoms within the unit cell have opposing polarizations for optic phonons near Γ.

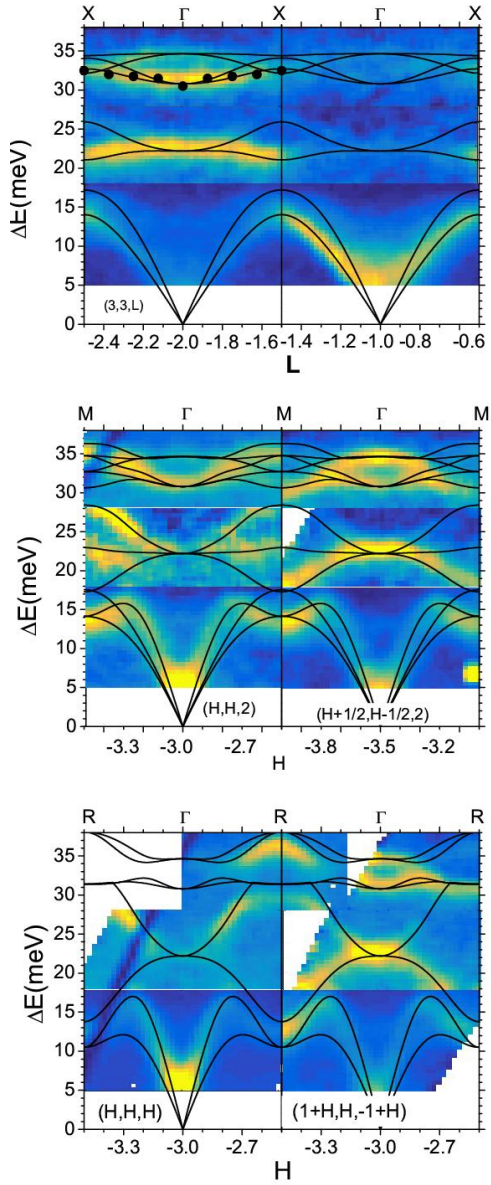

Figure 4: The measured phonon spectra along the (a) ΓX [3,3,L], (b) ΓM [H,H,2], and (c) ΓR.[H,H,H] and equivalent high symmetry directions, using an initial neutron energy $E_i$ = 60meV. The data are shown in different zones in the extended zone scheme, because the phonon structure factor is different in different zones. The solid lines are the result of the DFT phonon calculation for the case that the lattice constant is constrained to the experimental value and for the value $U = 6eV$; the energies obtained from the calculation have been multiplied uniformly by a factor 1.08 to give better agreement with the experimental data. The solid points in panel (a) represent the energy of the maximum in the magnetic resonance near 32meV as derived in Fig. 13; these clearly track the lower optic phonon branch at these energies.

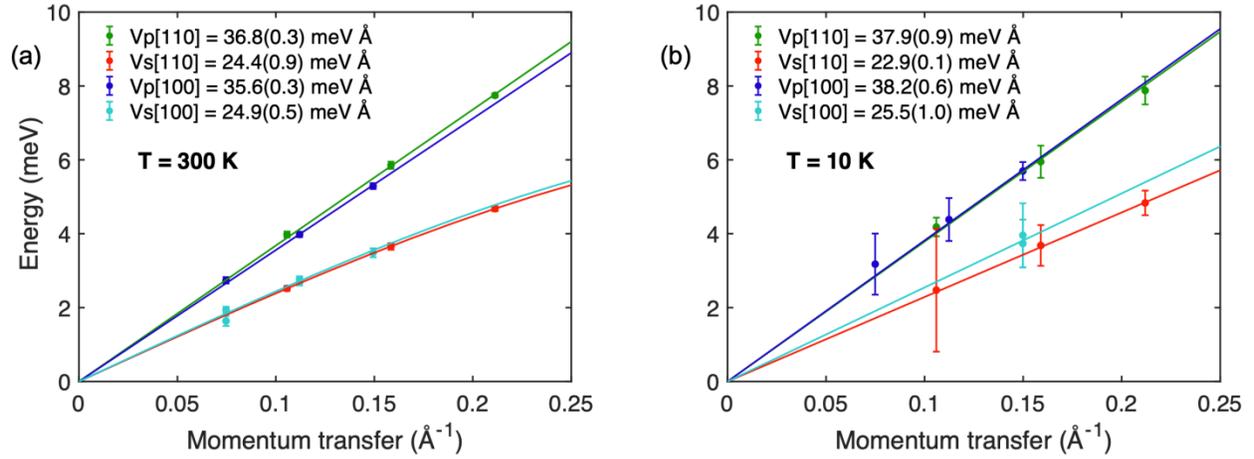

Fig. 5. Dispersion of acoustic phonons at the Brillouin zone centers of YbAl$_3$ at (a) 300K and (b) at 10K. The phonon velocities obtained from fits of linear or sine functions are indicated in the legend. V$_p$[HK0] and V$_s$[HK0] denote the velocities of longitudinal and transverse phonons, respectively, propagating along the [HK0] direction of reciprocal space.

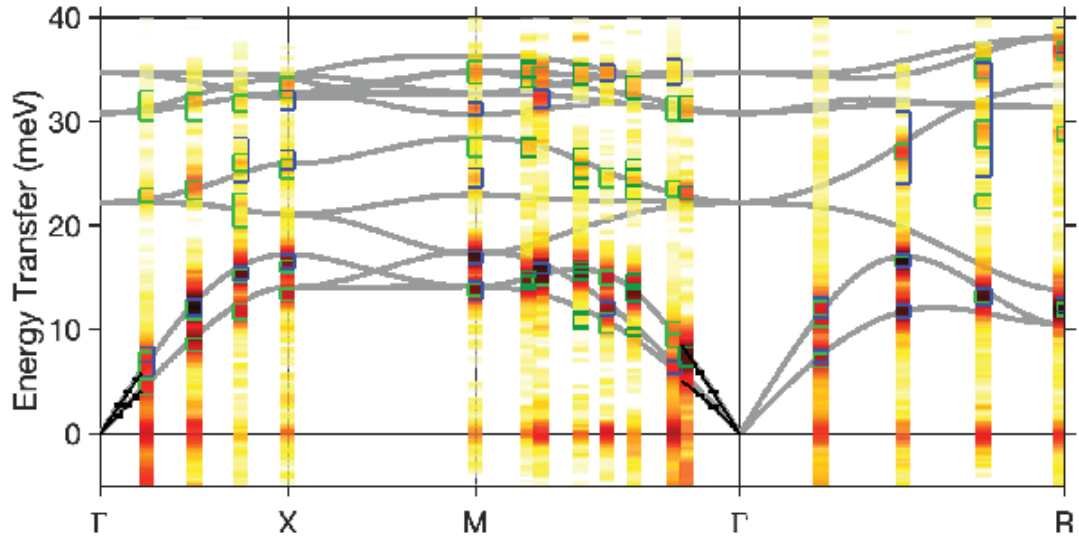

Fig. 6 Lattice fluctuation spectrum of YbAl$_3$ at 300 K as measured by inelastic x-ray scattering (IXS). Individual spectra are shown as color plots. The superimposed markers indicate phonon energies and linewidths determined by fits of the spectra. Blue and green markers denote phonons excited with an overall momentum transfer almost parallel to the [100] and [110] directions of reciprocal space, respectively. The calculated phonon bands of YbAl$_3$ are drawn as gray lines. The black lines and markers reproduce the data shown in Fig. 5.

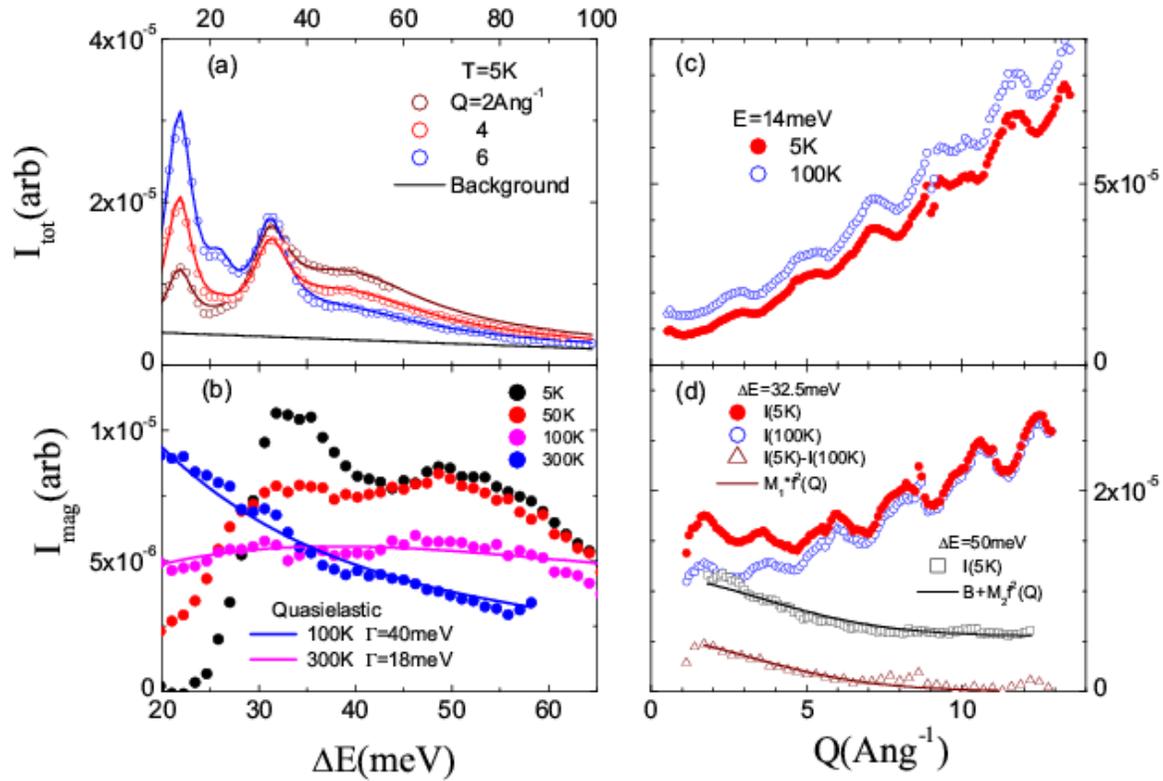

Figure 7: Inelastic neutron spectra for the polycrystalline average of the spectra observed for a single crystal of YbAl$_3$. The incident energy is 120meV. (a) The spectra for three values of momentum transfer **Q** at T=5K. The solid black line represents the background scattering. The solid colored lines represent fits to the sum of the background scattering, phonons at 14, 22, and 32meV, and two magnetic peaks, one at 32meV and one at 48meV. (b) The magnetic scattering obtained by subtracting the phonons and background as in panel (a) at four temperatures. The solid lines represent quasielastic scattering (with halfwidth $\Gamma$) which occurs at elevated temperature. (c) and (d) The **Q**-dependence of the scattering at select energies and at T=5 and 100K. The open triangles are the result of subtracting the 100K data from the 5K data for $\Delta E$=32.5meV. The solid line shows that this subtracted data follows the Yb 4$f$ form factor. The solid line for the $\Delta E$=50meV data is the sum of a constant background and a term tracking the 4$f$ form factor.

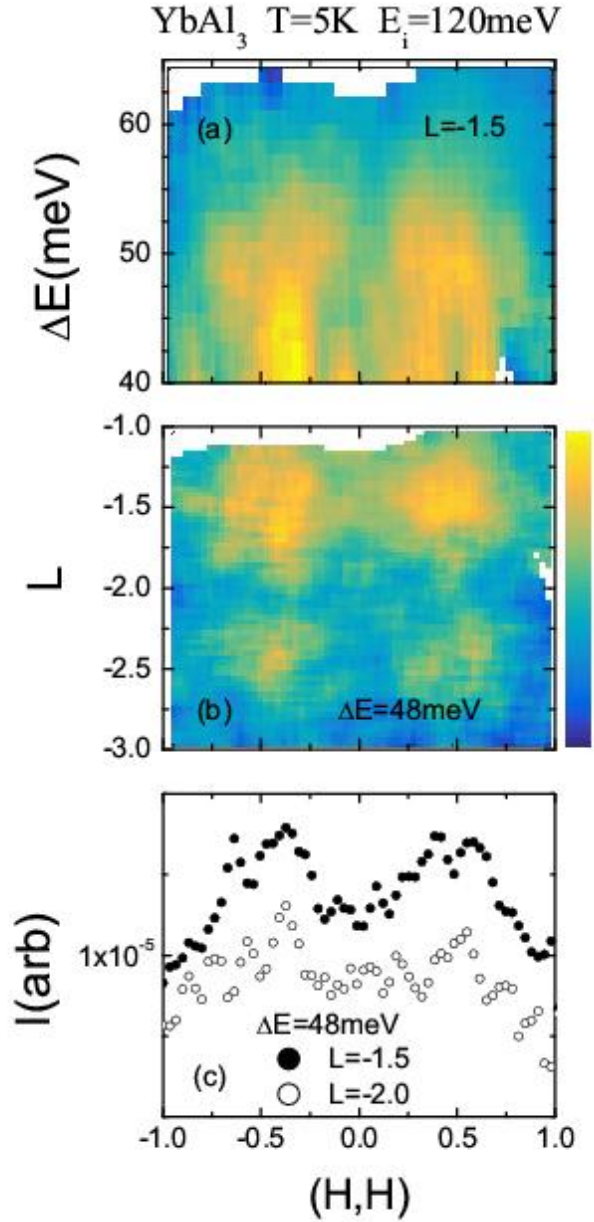

Figure 8: (a) A color **Q**-$\Delta E$ slice of the data along **Q**=(HH,3/2), showing columns of greater intensity for H ~ 0.35 and weak scattering for H=0. The same intensity variation can be seen in (b), which shows the scattering in the (HH,L) plane for $\Delta E$=48meV. Bright spots are seen near (0.35,0.35, 0.5) and symmetry-equivalent points, while weaker scattering is seen near (0,0,1.5) and yet weaker scattering near (0,0,2). (c) Cuts along (HH,3/2) and (HH,2) for $\Delta E$=48meV again show strong scattering near (0.35,0.35,1.5), weaker scattering near (0,0,1.5), and yet weaker scattering at (0,0,2).

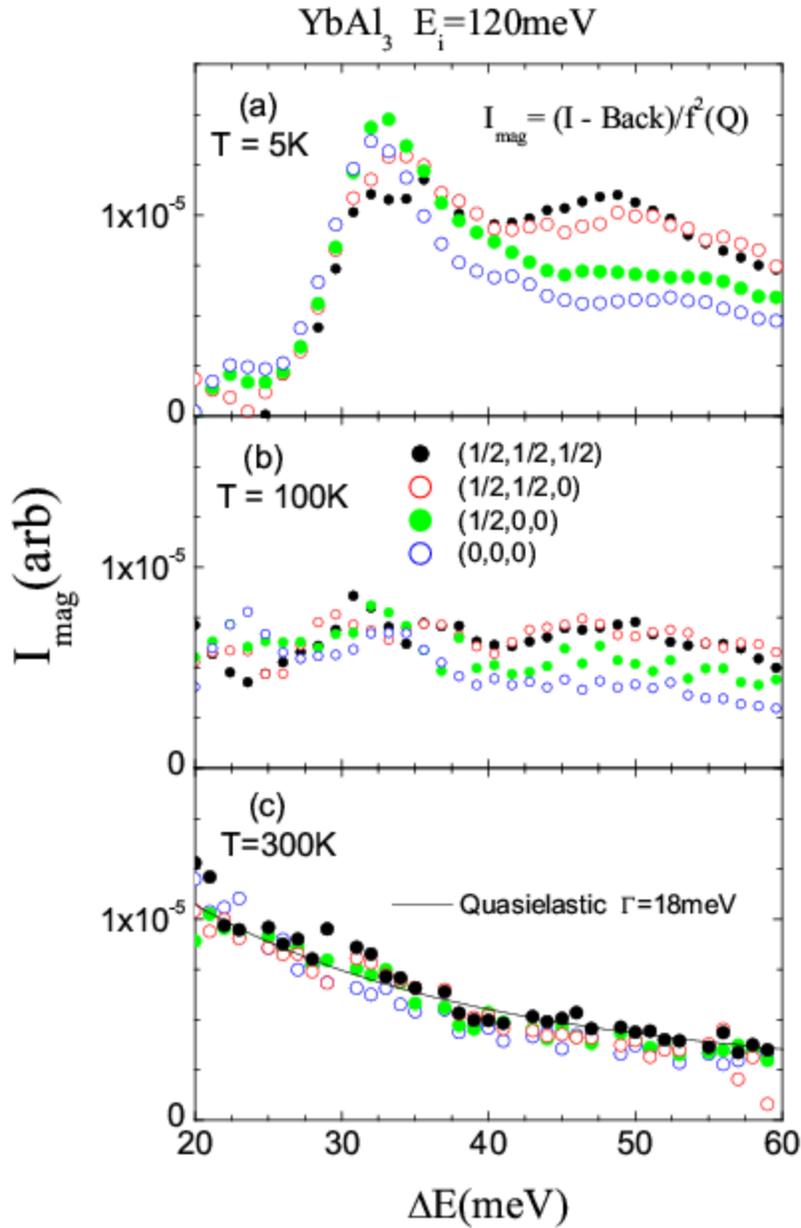

Figure 9: Magnetic spectra at three temperatures obtained from the total intensity by subtracting the background and dividing by the 4$f$ form factor to reduce the data to the first Brillouin zone. At 5K (panel (a)) the intensity in the Kondo scale (~48meV) peak decreases in the sequence I(1/2,1/2,1/2) > I (1/2,1/2,0) > I (1/2,0,0) > I(0,0,0) while the intensity in the peak near 32meV follows the opposite sequence. At 100K (panel (b)), the Kondo-scale scattering – albeit weaker – maintains this **Q**-dependence while the 32meV scattering is strongly reduced. At 300K (panel (c)) the scattering is **Q**-independent and quasielastic.

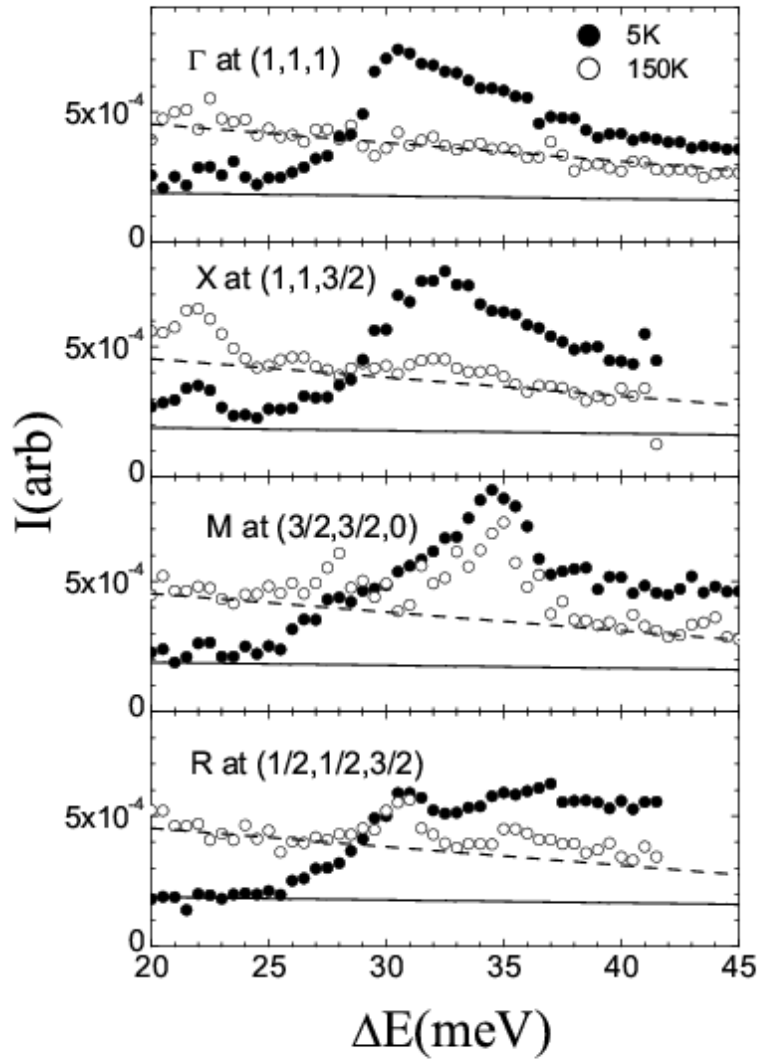

Figure 10: The measured total scattering at four values of **Q** and at 5 and 100K using an incident neutron energy $E_i$ = 60meV. The solid lines represent the background scattering; the dashed lines are the sum of the background and magnetic scattering at 100K. The 32meV phonon is essentially absent at (1,1,1) and (1,1,3/2), so that the magnetic scattering dominates, while the phonons dominate the 30-35meV spectra at (3/2,3/2,0) and (1/2,1/2,1/2).

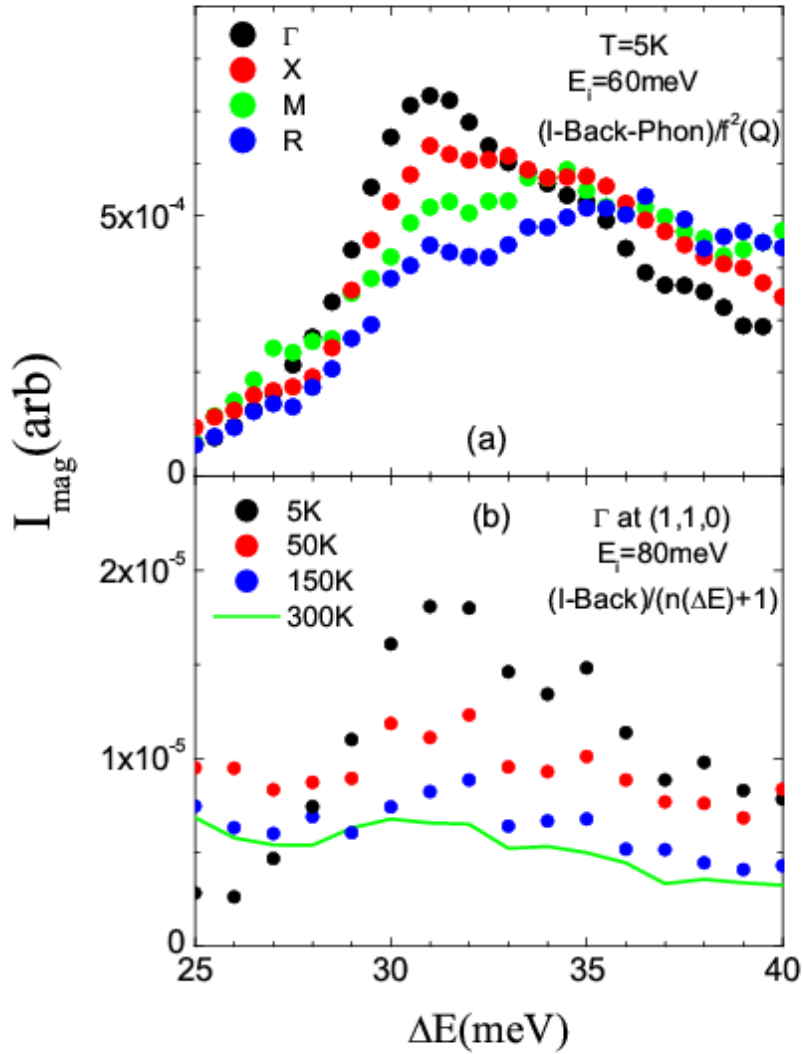

Figure 11: The behavior of the 32meV magnetic resonance. (a) The result of subtracting the background and the 30-35 meV optic phonons to obtain the magnetic spectra at 5K. The resonance is strong at $\Gamma$ and X, weaker at M, and vanishingly small at R where the Kondo scale scattering dominates. (b) The temperature dependence of the (1,1,0) magnetic scattering. The 32meV resonance vanishes rapidly as the temperature is raised.

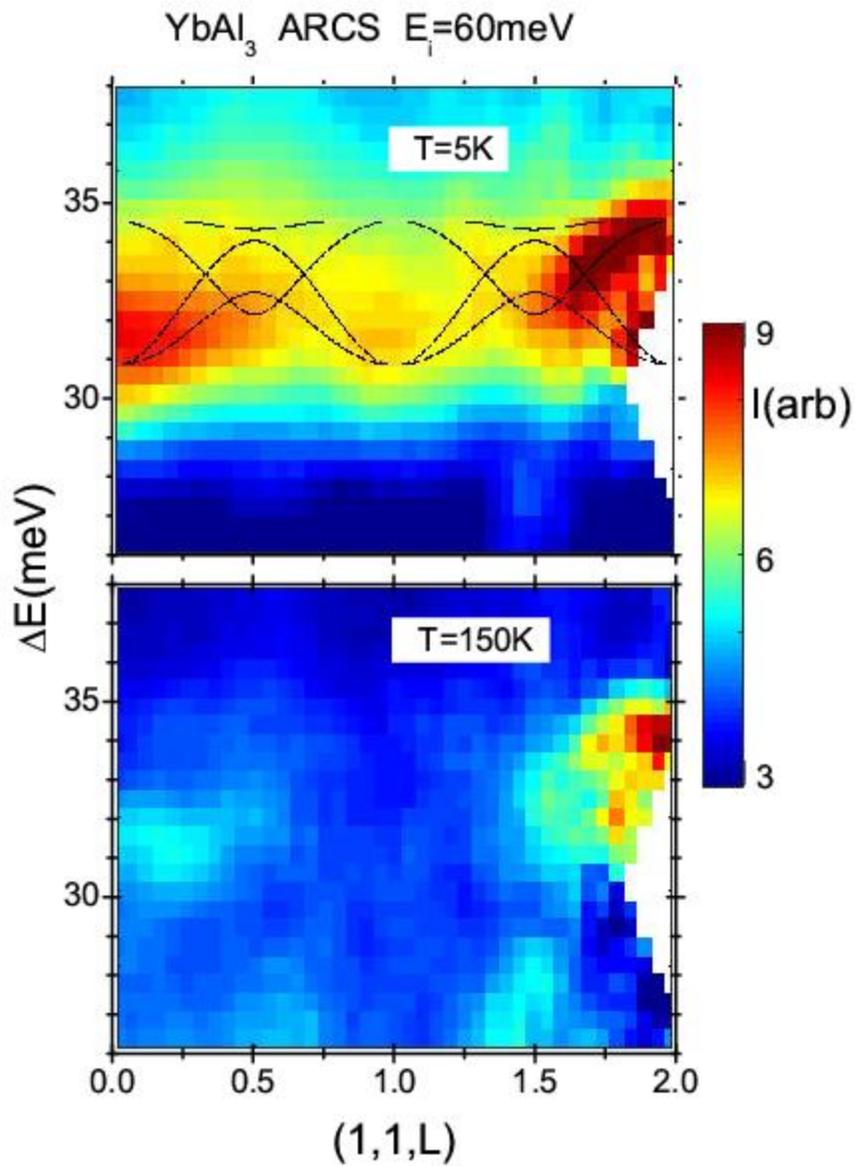

Figure 12 The top panel compares the resonant scattering along ΓX at 5K to the calculated phonons (with lattice constant constrained to the experimental value and with energies multiplied by an additional factor 1.08). The phonon structure factor vanishes at (1,1,1), so the scattering there is entirely magnetic. The leading edge of the resonance tracks the optic phonon dispersion along this direction. The lower panel shows that the resonance has disappeared by 150K.

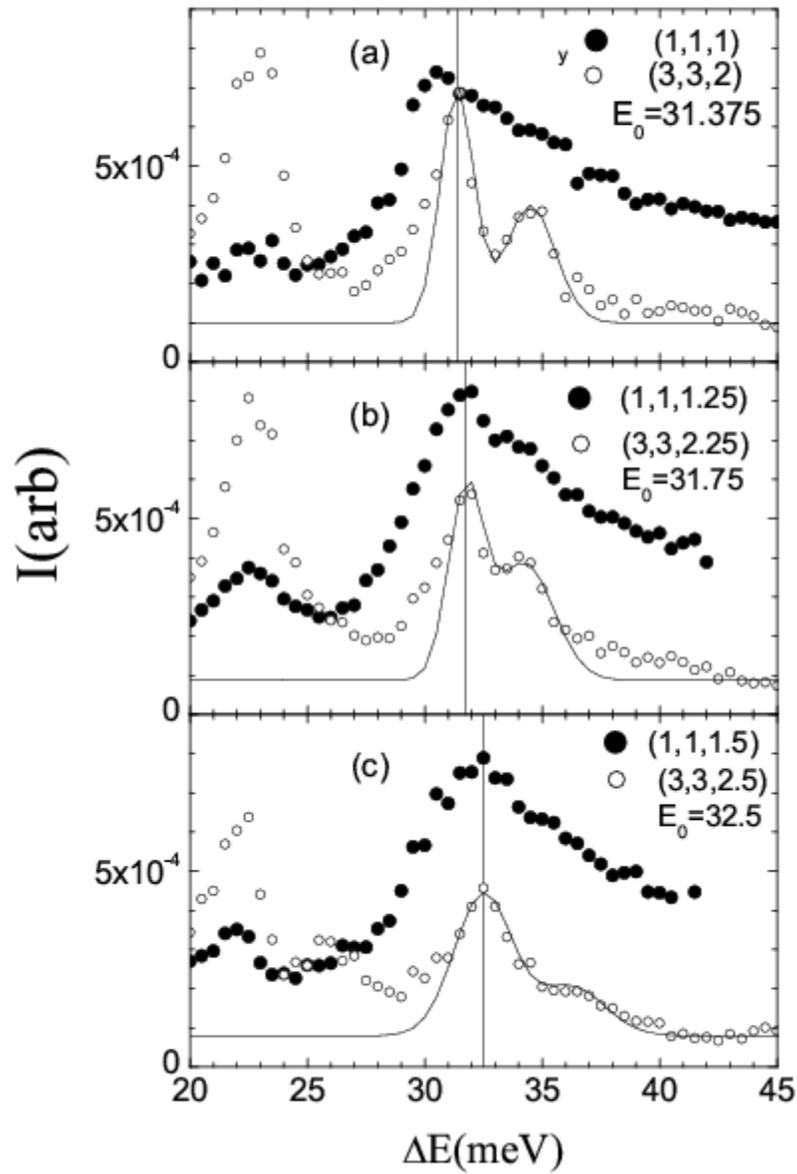

Figure 13: A comparison of the magnetic resonance-dominated scattering at 5K along (1,1,1+x) to the phonon-dominated scattering at (3,3,2+x). The initial neutron energy is $E_i = 60$meV. The magnetic resonance peak clearly tracks the increasing energy of the lower optic mode in the 30-35meV range.

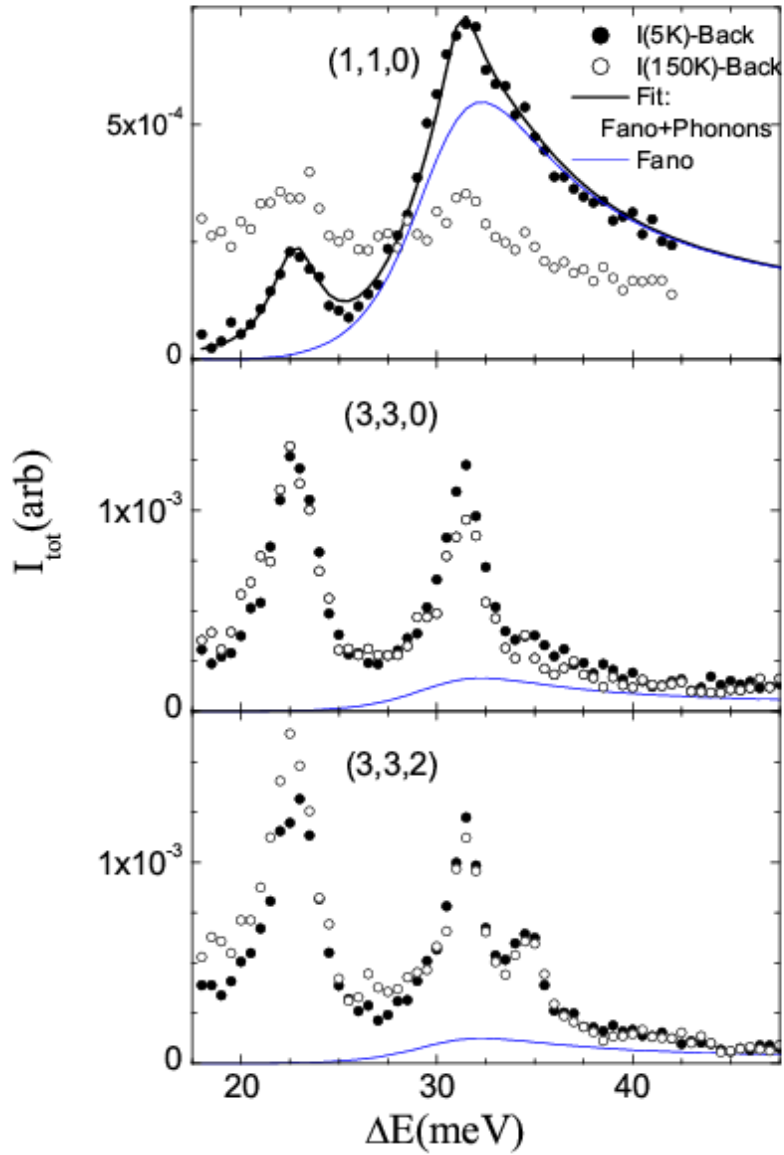

Figure 14: (a) The scattering minus background at (1,1,0) at 5K and 150K measured with an initial neutron energy $E_i = 60$meV. The phonon peaks at 150K have the same amplitude as at 5K. The solid line is the sum of these phonon peaks and a magnetic line simulated with a Fano lineshape (blue line) with the resonance factor q=2.5 and the bare linewidth 4.5meV. (b) The scattering at (3,3,0), is in an equivalent zone as (a). The magnetic contribution at 5K is reduced by the 4$f$ form factor. (c) The scattering at (3,3,2) is dominated by the phonons, with a small magnetic contribution at 5K. Note that the 31 and 35 meV peaks do not change position or amplitude with temperature.

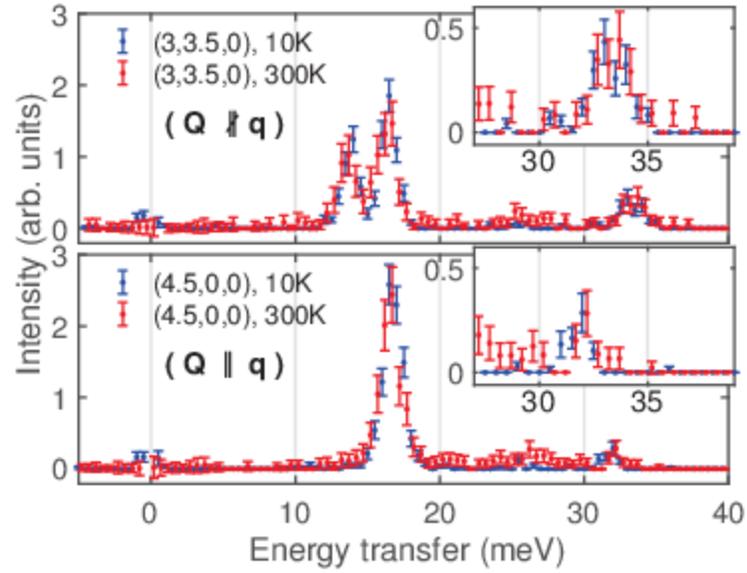

Figure 15: Comparison of IXS spectra at the X points of reciprocal space at 10K and 300K, in settings where **Q** is parallel (bottom panel) and not parallel (top panel) to $\mathbf{q}_{ph}$. The spectra have been corrected for the Bose population factor and an arbitrary scale factor has been applied.

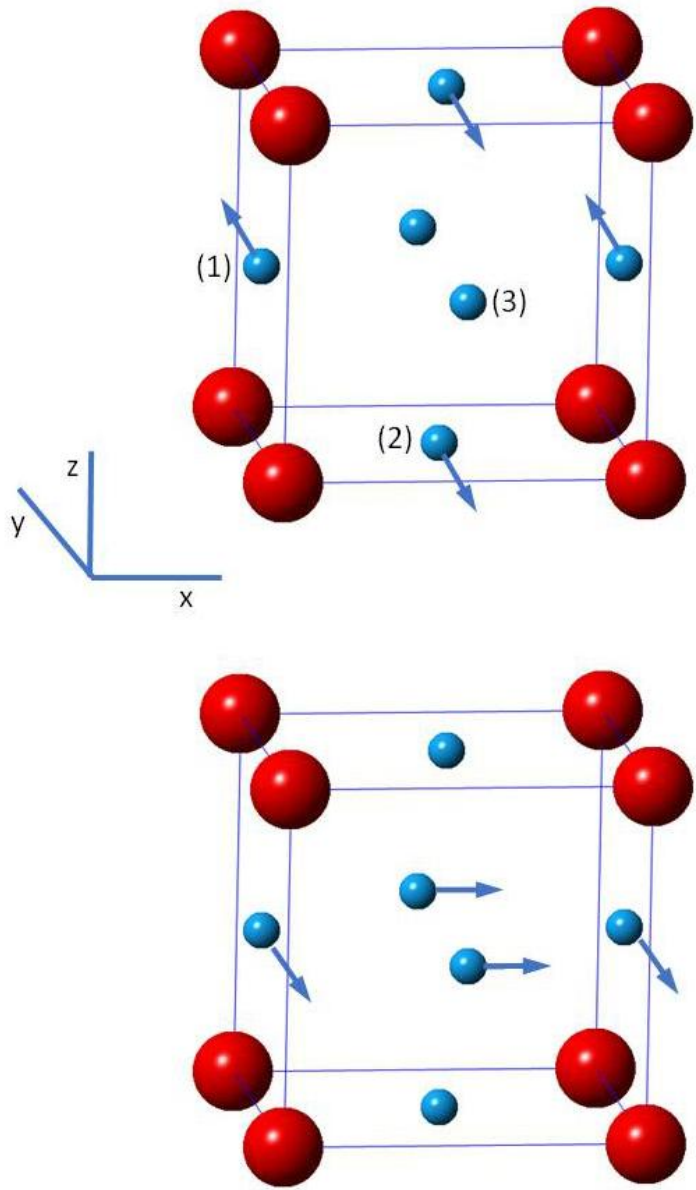

Figure 16: Two optic phonon oscillation modes near 32meV. Red spheres: Yb; blue spheres: Al. Upper panel: mode 9 at the Γ-point. In this mode, the Al atoms at (1/2,1/2,0) and (0,1/2) move with equal amplitude but in opposite directions along the y-direction, while the other Al atom and the Yb atom remain fixed. Lower panel: Mode 7 at the M-point. In this mode, the Al atoms at (1/2,0,1/2) move in the x-direction while the (0,1/2,1/2) atoms move in the –y direction with equal amplitude while the other atoms remain fixed. Mode 9 at Γ is associated with the resonance while mode 7 at M mode resonates weakly in comparison.